\documentclass[11pt, a4paper]{article}
\usepackage{jheppub}

\usepackage{euscript,xspace, color}
\usepackage{mathtools,rotating}



\title{Holographic Thermal Correlators for Hyperbolic $CFT$s}

\author[a]{Atanu Bhatta,}
\author[b]{Shankhadeep Chakrabortty,}
\author[c]{Taniya Mandal}
\author[d]{and Arpit Maurya}

\affiliation[a]{National Institute of Theoretical and Computational Sciences,
School of Physics and Mandelstam Institute for Theoretical Physics,
University of the Witwatersrand, Wits 2050, South Africa}
\affiliation[b, d]{Department of Physics, Indian Institute of Technology Ropar, Rupnagar, Punjab 140 001, India}
\affiliation[c]{Department of Physics, Indian Institute of Technology Kanpur,
Kanpur 208016, India}

\emailAdd{atanu.bhatta@wits.ac.za}
\emailAdd{s.chakrabortty@iitrpr.ac.in}
\emailAdd{taniyam@iitk.ac.in}
\emailAdd{arpit.20phz0009@iitrpr.ac.in}

\abstract{
We use holography to compute the exact form of retarded Green’s functions for a scalar operator with conformal dimension $\Delta$ in a thermal CFT and in its related counterpart with chemical potential in $R^1\times H^3$. In our analysis, we recast the wave equation of a scalar field in the normal form of Heun's equation in the dual gravity theories described by the AdS hyperbolic blackhole and its charged version. Heun's equation is identified to the  semiclassical limit of the BPZ equation for a five-point correlator with one degenerate field insertion in the Liouville theory on the Riemann sphere. The crossing symmetry of conformal block in the Liouville theory eventually gives rise to a set of connection formulas among the solutions of Heun's equation evaluated at different regular singularities. We use the connection formula to reproduce the leading order behaviors of the scalar field near the horizon as well as near the boundary and achieve the exact form of the retarded thermal Green's function. We show a recipe to obtain the exact 
 retarded Green's function for a thermal CFT dual to AdS blackbrane from a high-temperature limit accompanied by a complex mapping on AdS hyperbolic blackhole. Moreover, we show the retarded Green's function for the boundary CFT of Rindler AdS spacetime  admits a free integer parameter. 
}

\arxivnumber{2308.14704}

\begin{document}

\maketitle

\flushbottom

\section{Introduction}
A key achievement of the AdS/CFT correspondence \cite{Maldacena:1997re} is its ability to compute physical observables such as thermal correlation function for a strongly coupled thermal field theory by applying holographic techniques in a weakly coupled dual gravity theories. The prescription for computing thermal correlator by using the AdS/CFT correspondence was originally given in Euclidean signature\cite{Gubser:1998bc,Witten:1998qj,Witten:1998zw}. Further, the geodesic approximation method was conjectured to determine the thermal correlator of operators with large conformal dimensions by computing the geodesic action in the dual bulk theory \cite{Louko:2000tp,Rodriguez-Gomez:2021mkk,Rodriguez-Gomez:2021pfh,Georgiou:2022ekc}.

Various attempts have been made to extend Euclidean thermal correlators to the Lorentzian signature to explore real-time dynamics of strongly coupled thermal field theory\cite{Balasubramanian:1998de,Balasubramanian:1999ri,Balasubramanian:1998sn}. However, there are subtleties in executing such analytical continuation for thermal correlators.  In particular, Lorentzian continuation can be successfully addressed once the exact form of the Euclidean correlator is available for all Matsubara frequencies whereas, on the contrary, the holographic computation of thermal correlator in dual AdS blackhole spacetime is achievable only under certain approximations, e.g., the low and high-temperature limits \cite{Policastro:2001yb}. Subtleties of analytical continuation were avoided by introducing a direct approach to compute the thermal correlator in Lorentzian signature by generalizing the holographic prescription in Euclidean AdS/CFT to the one in Minkowskian AdS/CFT \cite{Son:2002sd}. According to the Miskowskian prescription, the wave equation of a scalar field in an $AdS$ blackhole background admits a mode expansion that satisfies an incoming boundary condition near the horizon. The boundary limit of such a solution is given by a linear combination of normalizable and non-normalizable modes interpreted as a response function and a source term respectively. Now, the retarded Green's function becomes a ratio of the response function to the source term. Finally, we  identify the momentum-space Matsubara correlator which is essentially the Fourier-transformed version of the boundary thermal correlator, to the retarded Green's function. The real-time thermal correlators are very crucial to study the near-equilibrium behavior \cite{Policastro:2001yc}, transport properties \cite{Hartnoll:2016apf} and the chaotic behaviour \cite{Grozdanov:2017ajz,Blake:2018leo} of the strongly coupled thermal field theory.

Although some exact results for the thermal correlator in the finite temperature field theory dual to the thermal AdS spacetime have been achieved within the holographic framework \cite{Alday:2020eua}, such analysis in the thermal field theory of general dimension dual to AdS blackhole has been out of reach until recently. In \cite{Dodelson:2022yvn} the authors have presented an exact result in four-dimensional thermal CFT dual to $AdS_5$ Schwarzschild blackhole by using some appropriate techniques originally developed in super-symmetric gauge theory \cite{Seiberg:1994rs, Seiberg:1994aj, Nekrasov:2002qd, Alday:2009aq, Nekrasov:2009rc}. To find the thermal correlator the wave equations are solved on the black hole background via Nekrasov-Shatashvili functions~\cite{Aminov:2020yma}. However these wave equations can be identified as the BPZ equation in Liouville theory on the Riemann sphere as well \cite{Musiri:2003rs, BarraganAmado:2018zpa,Amado:2021erf}.

More specifically, the BPZ equation for the five-point correlator with an insertion of the degenerate operator in Liouville 
theory allows series solutions in terms of conformal blocks \cite{Piatek:2017fyn,Dolan:2000ut,Dolan:2003hv,Dolan:2011dv}. The crossing symmetry of the conformal blocks evaluated in various OPE channels, e.g., $s$, $t$ and $u$ channels  establishes crossing relations  among the corresponding solutions to the BPZ equation.  In the large central charge limit (semi-classical limit), the BPZ equation takes the form of  Heun's equation  \cite{Bonelli:2022ten}, a second-order linear differential equation with four regular singularities \cite{fiziev2015heun,ronveaux1995heun}. In the context of Liouville CFT, those singular points correspond to primary operator insertions in the  theory. The crossing relation among the solutions of the semi-classical BPZ equation induces a connection formula among the solutions of Heun’s equation. Such connection formula  essentially tells that a solution of the Heun's equation also known as the Heun function 
near one singular point is expandable in terms of the solutions around other
singularities \cite{Bonelli:2022ten}. A detailed description of the connection formula is given in \cite{Lisovyy_2022}.  

Such connection formula plays a crucial role in understanding Alday-Gaiotto-Tachikawa (AGT) correspondence according to which the BPS sector of four-dimensional $\mathcal{N} =2$, $SU(2)$ gauge theory in $\Omega$ background turns out to be dual to the Liouville CFT on Riemann sphere \cite{Alday:2009aq,Alday:2009fs}. As a consequence of the duality, the four-point function of the Liouville CFT maps to the Nekrasov partition function in $\mathcal{N} =2$, $SU(2)$ gauge theory \cite{Maruyoshi:2010iu}. Further, the semi-classical limit on the parameters of Liouville CFT identifies with the  Nekrasov-Shatashvili limit on the parameters of the $\Omega$ background in the $SU(2)$ gauge theory \cite{Nekrasov:2009rc}. This connection has sparked renewed interest in the study of Heun's equation and its applications in diverse fields \cite{Hortacsu:2018rr, Kwon:2023ghu, He:2023wcs}. A leading non-perturbative correction to a two-sided thermal correlator is discussed using the connection formula \cite{Dodelson:2023vrw}. The bulk computation of the $HHLL$ correlator heavily relies on similar techniques \cite{Giusto:2023awo}. The connection formula plays a crucial role in understanding the linear perturbation theory of blackhole spacetime and its associated quasinormal modes \cite{Aminov:2023jve} and the references therein.

In \cite{Dodelson:2022yvn}, the authors derive a connection formula between the Heun functions associated with the incoming mode expansion of a scalar field near the horizon and the other near the boundary of the $AdS_5$ Schwarzschild blackhole.
Further, they show that the same connection formula determines the response function and source term, and therefore it is possible to obtain an exact retarded Green's function for a thermal CFT living in $R^1 \times S^3$. Some of the authors in the present manuscript computed the exact form of the retarded Green's function for thermal $CFT_4$ with chemical
potential and angular momenta on $R^1\times S^3$\cite{Bhatta:2022wga}.

Generalizing the analysis presented in \cite{Dodelson:2022yvn, Bhatta:2022wga}, in this paper, we compute an exact form of  retarded Green's function for a thermal CFT living in $R^1\times H^3$. 
In particular, we study massive scalar mode in the dual $AdS_5$ hyperbolic background and derive the normal form of Heun's equation. Finally, we achieve an exact form of thermal retarded Green's function from the connection formula associated with the respective Heun function. 
Also, we generalize our result for a charged $AdS_5$ hyperbolic background dual to a thermal CFT with chemical potential living in $R^1\times H^3$. In \cite{Dodelson:2022yvn}, the authors have shown how to obtain the connection formula for $AdS_5$ blackbrane from a high-temperature limit on $AdS_5$ Schwarzschild blackhole. We show that the appropriate connection formula for $AdS_5$ blackbrane can be reproduced by taking a high-temperature limit accompanied by a complex mapping in $AdS_5$ hyperbolic background. Finally, we show that the exact form of retarded Green's function in Rindler $AdS_5$ background curiously depends on a free integer parameter. 

The plan of the paper is the following: In section \ref{sec:hypbh}, we discuss the computation of the exact form of retarded Green's function for the CFT in $R^1\times H^3$. In section \ref{sec:hypbh1}, we generalize our analysis for a CFT with chemical potential in $R^1\times H^3$ followed by the computation of the same for a thermal CFT in $R^1 \times R^3$ in section \ref{sec:bb}. Exact thermal correlators for Rindler $AdS_5$ have been analyzed in section \ref{sec:rindler}. Finally, we conclude in section \ref{sec:conclusion}.

\section{Holographic computation of exact retarded Green's function for thermal CFT in $R^1 \times H^3$}\label{sec:hypbh}
Unlike the flat spacetime ($\Lambda = 0$) and the de Sitter (dS) spacetime ($\Lambda >0$), the Anti-de Sitter (AdS) spacetime (
$\Lambda<0$) admits solutions of Einstein's equation described as the hyperbolic AdS blackhole \cite {Birmingham:1998nr, Emparan:1998he, Emparan:1999gf, Ahn:2019rnq}. AdS hyperbolic Reissner-Nordstrom black holes and AdS hyperbolic blackhole in gauged supergravities are also available in the literature \cite{Cai:2004pz,Hosseini:2019and, Svesko:2020dfw,  He:2010zb,Huang:2022avq}. Consequently, by the virtue of generalized $AdS_5/CFT_4$ correspondence, one can holographically associate the thermal states of CFT living on the $3+1$ dimensional hyperbolic boundary  ($R^1 \times H^3$) of the blackhole geometry \cite{Hung:2011nu, Casini:2011kv}. In this section, by following \cite{Dodelson:2022yvn}, we derive the exact form of retarded Green's function for the boundary CFT living in $R^1 \times H^3$. Later, we  generalize our result for the hyperbolic CFT$_4$ in the presence of a chemical potential. 

\label{uncharged}

The Einstein-Hilbert action with a negative cosmological constant in five dimensions is given by 
\begin{equation}
    S=\frac{1}{16\pi G_5}\int d^5x \sqrt{-g}\left(R+\frac{12}{L^2}\right),
\end{equation}
where $G_5$ is the five-dimensional Newton's constant and $L$ is the radius of curvature of AdS spacetime. The equation of motion derived from the above action admits a solution of the form of $4+1$ dimensional hyperbolic black hole 

\begin{equation} \label{metric1}
d s_{HAdS_5}^2=-f(r) d t^2+\frac{d r^2}{f(r)}+r^2 d H_3^2.
\end{equation}
The blackening factor $f(r)$ is
\begin{equation}
 f(r)=-1-\frac{M}{r^2}+\frac{r^2}{L^2},
 \label{hypbh}
\end{equation}
where $M$ is related to the $ADM$ mass of the blackhole. It is evident from eqn (\ref{hypbh}) that the extrinsic curvature of the blackhole horizon is $-1$ and therefore the corresponding topology is hyperbolic ($R^1 \times H^3$).  We consider the boundary topology also as $R^1 \times H^3$ where the explicit form of the corresponding spatial section of the boundary metric takes the form 
\begin{equation}
\label{AdShyp}
 d H_3^2=d \chi^2+\sinh ^2 \chi \left(d \theta^2+\sin ^2 \theta d \phi^2\right).
\end{equation}
For computational convenience, the blackening factor $f(r)$ is re-expressed as follows
\begin{equation}
f(r)=\left(1-\frac{r_h^2}{r^2}\right)\left(-1+\frac{r^2}{L^2}+\frac{r_h^2}{L^2}\right),
\end{equation}
where the radial position of the horizon becomes
\begin{equation}
r_h = \sqrt{\frac{\sqrt{L^4 + 4 M L^2} + L^2}{2}}
\label{hyphor}
\end{equation}
The Hawking temperature  of the black hole is given by  \cite{Emparan:1999gf}
\begin{equation} \label{Thyp}
    T = 
    \frac{1}{\pi}\left(\frac{r_h}{L^2}-\frac{1}{2r_h}\right)
\end{equation} 
and it is holographically identified to the temperature of the thermal state of the boundary CFT. 

In order to facilitate our  analysis, we consider a scalar field theory in the hyperbolic black hole background such that the mass of the scalar field $\phi$ is holographically related to the conformal dimension $\Delta$ of the primary operator $\mathcal{O}$ in the boundary CFT living in $R^1 \times H^3$ 
\begin{equation}
m L=\sqrt{\Delta(\Delta-4)}.
\end{equation}

The Klein-Gordon equation satisfied by the scalar field $\phi$ takes the form,
\begin{equation}\label{KGeq}
\left(-\frac{1}{f(r)}\frac{\partial^2}{\partial t^2} + \frac{1}{r^3}\frac{\partial}{\partial r}  \left( r^3 f(r) \frac{\partial}{\partial r}\right) + \frac{1}{r^2} \Box_{H_{3}}-m^2\right) \phi=0 
\end{equation} 
where $\Box_{H_{3}}$ is the Laplacian of the 3-dimensional hyperbolic space.\\
Assuming the time translation symmetry we choose an ansatz for the scalar field in terms of its Fourier modes, 
\begin{equation}\label{fourier}
\phi(t,r,\chi,\Omega_2)=\int d \omega \int d \lambda \sum_{\Vec\mu} e^{-i \omega t} \Psi_{\omega \lambda }(r) Y_{\lambda {\Vec\mu}}\left(\chi, \Omega_2\right)
\end{equation}
where $\Psi_{\omega \lambda }(r)$ is the radial mode function and $Y_{\lambda {\Vec\mu}}\left(\chi, \Omega_2\right)$ is the hyperbolic spherical harmonic satisfying the eigenvalue equation on $H^3$\cite{analyticmap,David:2022nfn},
\begin{equation}\label{hyperharmonic}
\square_{H_{3}} Y_{\lambda_ {\Vec\mu}}\left(\chi, \Omega_{2}\right)=-\left(\lambda^2 +1\right) Y_{\lambda  {\Vec\mu}}\left(\chi, \Omega_{2}\right)
\end{equation}
Inserting  \eqref{fourier} in to \eqref{KGeq} and by using \eqref{hyperharmonic}, the radial equation reads, 
\begin{equation}
\left(\frac{1}{r^3} \partial_r\left(r^3 f(r) \partial_r\right)+\frac{\omega^2}{f(r)}-\frac{\left(\lambda^2+1\right)}{r^2}-\frac{\Delta(\Delta-4)}{L^2}\right) \Psi_{\omega \lambda}(r)=0
\label{radial}
\end{equation}
In order to perform the computation of the retarded Green's function we need to solve the wave equation with appropriate boundary conditions. Here we impose ingoing boundary conditions on the horizon and find
\begin{equation} \label{rradeqhads}
\Psi_{\omega \lambda}^{\text {in }}(r)=\left(r-r_{h}\right)^{-\frac{i \omega}{2} \frac{r_{h}}{\left(\frac{2 r_{h}^2}{L^2}-1\right)}}+\ldots
\end{equation}
whereas the solution near the AdS boundary $r\rightarrow \infty$ behaves as 
\begin{equation}
\Psi_{\omega \lambda}^{\mathrm{in}}(r)=\mathcal{A}(\omega, \lambda)\left(r^{\Delta-4}+\ldots\right)+\mathcal{B}(\omega, \lambda)\left(r^{-\Delta}+\ldots\right) .
\end{equation}
The ratio of the response $\mathcal{B}(\omega, \lambda)$ to the source  $\mathcal{A}(\omega, \lambda)$ is regarded as the retarded Green's function \cite{Son:2002sd},
\begin{equation} \label{rtgrnfn}
G_R(\omega,\lambda)=\frac{\mathcal{B}(\omega, \lambda)}{\mathcal{A}(\omega,\lambda)}
\end{equation}
In order to bring the equation \eqref{radial} to a form of Heun's equation, we introduce a set of transformations in the radial coordinate $r$ and the radial mode function $\Psi_{\omega \lambda }(r)$,
\begin{equation} \label{trans11}
z =\frac{r^2}{r^2+r_{h}^2-L^2}
\end{equation}
\begin{equation}  \label{trans2}
\Psi_{\omega \lambda}(r)  =\left(r^3 f(r) \frac{d z}{d r}\right)^{-1 / 2} X_{\omega \lambda}(z(r))
\end{equation}
We can invert the field redefinition 
\eqref{trans2} and express $X_{\omega \lambda}$ in terms of only $z$,
\begin{equation}
\label{inverse}
    X_{\omega \lambda}^{in}(z)= \left.\alpha(r) \Psi_{\omega\lambda}^{in}(r) \right\vert_{r=\frac{\sqrt{r_h^2-L^2}}{\sqrt{1-z}}z}, 
\end{equation}
where 
\begin{equation}
\begin{aligned}
    \alpha(r) & =  \left(r^3 f(r) \frac{d z}{d r}\right)^{1 / 2} = \left( \frac{2r^2}{L^2} \frac{(r^2-r_h^2)(r_h^2-L^2)}{(r^2+r_h^2-L^2)} \right)^{1/2}.
\end{aligned}
\end{equation}
 In terms of the newly defined entities $z$ and $X_{\omega \lambda}(z)$, the radial equation \eqref{radial} is transformed into  
 Heun's differential equation in normal form,
 \begin{eqnarray}
\label{hypheun}
\left(\partial_z^2+\frac{\frac{1}{4}-a_1^2}{(z-1)^2}-\frac{\frac{1}{2}-a_0^2-a_1^2-a_t^2+a_\infty^2+u}{z(z-1)}+ \frac{\frac{1}{4}-a_t^2}{(z-t)^2} + \frac{u}{z(z-t)}+ \frac{\frac{1}{4}-a_0^2}{z^2} \right) X_{\omega \lambda}(z) = 0
 \end{eqnarray}
It is straightforward to 
express the parameters of the Heun's equation in terms of the parameters of the hyperbolic blackhole spacetime,

\begin{equation}\label{hyppara1}
\begin{array}{|c|c|c|c|c|c|}
\hline \text { Heun's Parameter } & t & a_0 & a_t & a_1 & a_{\infty} \\
\hline \text {AdS hyperbolic black hole } & \frac{r_{h}^2}{2 r_{h}^2-L^2} & 0 & \frac{i \omega}{2} \frac{r_{h}L^2}{2 r_{h}^2-L^2} & \frac{\Delta-2}{2} & \frac{\omega L^2}{2} \frac{\sqrt{r_{h}^2-L^2}}{2 r_{h}^2-L^2} \\
\hline
\end{array}
\end{equation}
with $u$ as
\begin{equation}
\begin{aligned}
u=  -\frac{(\lambda^{2} +1)L^2+2\left(2 r_{h}^2-L^2\right)+r_{h}^2 \Delta(\Delta-4)}{4\left(r_{h}^2-L^2\right)} 
 +\frac{r_{h}^2 L^4}{(r_{h}^2-L^2)} \frac{\omega^2}{4\left(2 r_{h}^2-L^2\right)} .
\end{aligned}
\label{u}
\end{equation}
Note that the Heun's equation \eqref{hypheun} exhibits four regular singular points located at $z =0, t, 1$ and $\infty$ among which 
$z=t$ and  $z=1$ correspond to the horizon and the boundary respectively. It is important to mention that due to the presence of a relative negative sign in the denominator of the r.h.s. of \eqref{trans11}, $z$ remains a well-defined real positive quantity only when the dual gravity is described as a large hyperbolic blackhole ($r_h>>L$). 

 Heun's equation \eqref{hypheun} near the black hole horizon $z=t$ becomes
\begin{equation}
\label{heunhor}
\left(\partial_z^2+\frac{\frac{1}{4}-a_t^2}{(z-t)^2}\right)X_{\omega\ell}(z)=0,
\end{equation}
By imposing the in-going  boundary condition, the leading behavior of the solution takes the form as , 
\begin{equation} \label{zradeqhads}
X_{\omega \lambda}^{\text {(t),in }}(z)=(t-z)^{\frac{1}{2}-a_t}+\ldots 
\end{equation}
Similarly, Heun's equation \eqref{hypheun} near the boundary ($z=1$) becomes
\begin{equation}\left(\partial_z^2+\frac{\frac{1}{4}-a_1^2}{(z-1)^2}\right)X_{\omega\ell}(z)=0.
\end{equation}
and the leading behaviour of a general solution takes the form,
\begin{equation} \label{chiexp0}
\begin{aligned}
X_{\omega \lambda}^{\text {(1) }}(z) & \propto 
\mathcal{A}(\omega,\lambda)X_{\omega \lambda}^{\text {(1,+) }} + \mathcal{B}(\omega,\lambda)X_{\omega \lambda}^{\text {(1,-) }},
\end{aligned}
\end{equation}
where we consider, 
\begin{eqnarray} \label{chiexp1}
X_{\omega \lambda}^{\text {(1,-) }} = \left(\frac{1-z}{r_{h}^2-L^2}\right)^{\frac{1}{2}-a_1}+\ldots ~~~ \& ~~~
X_{\omega \lambda}^{\text {(1,+) }} =
\left(\frac{1-z}{r_{h}^2-L^2}\right)^{\frac{1}{2}+a_1}+\ldots 
\end{eqnarray}
It is important to highlight that the appearance of the constant factor $(r_h^2 - L^2)$ in the denominator of the leading behaviour of $\chi_{\omega \lambda}^{\text {(1,-) }}$ and $\chi_{\omega \lambda}^{\text {(1,+) }}$ can be better understood if we take boundary limit in \eqref{inverse}.

\subsection{Relation to Liouville CFT}
We find the Heun's equations to appear in the study of conformal blocks in Liouville CFT. In this theory, the representation of Virasoro algebra contains degenerate operators which are essentially primary operators with null descendants. We introduce a parameter $b$ such that the central charge $c$ of the CFT can be written as
\begin{eqnarray}
    c = 1+ 6 \left( b + \frac{1}{b} \right)^2
\end{eqnarray}
Then the simplest example of a null descendant in the level two is
\begin{eqnarray}
    \left(L_{-1}^2 + b^2 L_{-2}\right)|h_{1,2}\rangle 
\end{eqnarray}
whereas the holomorphic conformal dimension of the corresponding degenerate primary operator $\Phi$ is $h_{1,2} = - \frac{1}{2} - \frac{3}{4b^2}$. In general, the degenerate primary operators are determined by the holomorphic conformal dimensions which satisfy the Kac formula
\begin{eqnarray}
    h_{r,s} = \frac{b^2}{4}(1-r^2) + \frac{1}{4b^2} (1-s^2) + \frac{1}{2}(1-rs)
\end{eqnarray}

Let us now consider the five-point function with one degenerate operator insertion
\begin{eqnarray}
  \mathcal{F}(z) = \langle 0|\mathcal{O}_{\infty}(\infty) \mathcal{O}_1(1) \mathcal{O}_t(t) \Phi(z) \mathcal{O}_0(0)|0\rangle 
\end{eqnarray}
where $\mathcal{O}_{\infty},\, \mathcal{O}_1,\, \mathcal{O}_t$ and $\mathcal{O}_0$ are the primary operators with the scaling dimensions $\Delta_{\infty},\, \Delta_1,\, \Delta_t$ and $\Delta_0$ respectively. The five-point function satisfies the following BPZ equation
\begin{eqnarray}
    && \left( b^{-2}\partial_z^2 + \frac{\Delta_1}{(z-1)^2}- \frac{\Delta_1+t \partial_t + \Delta_t+z \partial_z + \Delta_{2,1}+\Delta_0-\Delta_{\infty}}{z(z-1)} \right. \cr 
    && \qquad \qquad \qquad \qquad \left. + \frac{\Delta_t}{(z-t)^2}+ \frac{t}{z(z-t)}\partial_t - \frac{1}{z}\partial_z + \frac{\Delta_0}{z^2}\right) \mathcal{F}(z) = 0
\end{eqnarray}

The conformal correlators enjoy the conformal block expansions. For example, the five-point function can be written as
\begin{eqnarray}
    \mathcal{F}(z) =\sum_{\Delta,\,\ell} \sum_{\Delta',\,\ell'} \sum_a \lambda_{\Delta_{\infty}\Delta_1(\Delta, \ell)}\lambda^a_{(\Delta,\ell)\Delta_t(\Delta',\ell')} \lambda_{(\Delta',\ell')\Delta_{1,2}\Delta_0} W^{(a)}_{(\Delta,\ell) (\Delta',\ell')}(z)
\end{eqnarray}
with
\begin{eqnarray}
    W^{(a)}_{(\Delta,\ell) (\Delta',\ell')}(z) = P(z)\, G^{(a)}_{(\Delta,\ell) (\Delta',\ell')}(z)
\end{eqnarray}
where $\lambda_{\Delta_{\infty}\Delta_1(\Delta, \ell)}$ and $\lambda_{(\Delta',\ell')\Delta_{1,2}\Delta_0}$ are the OPE coefficients, $\lambda^a_{(\Delta,\ell)\Delta_t(\Delta',\ell')}$ is an independent coefficient whereas  $a$ stands for conformal family and $G^{(a)}_{(\Delta,\ell) (\Delta',\ell')}(z)$ is the five-point conformal block. The constant coefficients depend on the dynamical contents of the theory. The BPZ equation satisfied by the five-point conformal block is a second order ODE with four regular singular points at $z=0,1,t$ and $\infty$. The solutions of the ODE around different singular points correspond to the conformal blocks that appear in the expansions of the five-point correlation function in different channels. The five-point conformal blocks associated with different channels are related to each other by crossing relations.

In the semi-classical limit
\begin{eqnarray}
    b \to 0 \quad \text{ and } \quad \alpha_i \to 0 \quad \text{ such that } \quad b \alpha_i = a_i \text{ (finite) }
\end{eqnarray}
where
\begin{eqnarray}
    \Delta_i = \alpha_i (Q-\alpha_i); \qquad Q = \left( b + \frac{1}{b}\right)
\end{eqnarray}
the BPZ equation reduces to the Heun's equation in normal form. So, the semi-classical conformal blocks associated with the different channels are the solutions of the Heun's equation near different singular points, namely $z=0,1,t$ and $\infty$. The solutions corresponding to the different singular points are related to each other via crossing relations and these give rise to the various connection formulae as listed in~\cite{Bonelli:2022ten}.

For example, the connection formula between the solution $\chi^{(t)}$ near $z=t$ and the solution $\chi^{(1)}$ near $z=1$ reads
\begin{eqnarray}
\label{eq:connection}
\chi^{(t)}(z) &=& \sum_{\theta'=\pm} \left(\sum_{\theta=\pm} \mathcal{M}_{-\theta}(a_{t}, a; a_0)\mathcal{M}_{(-\theta)\theta'}(a, a_1; a_{\infty}) \right. \cr  
&& \qquad \qquad \left. t^{\theta a}e^{-\frac{\theta}{2}\partial_a F} \right) t^{\frac{1}{2}-a_0-a_{t}}(1-t)^{a_t-a_1} e^{-\frac{1}{2}(\partial_{a_t}+\theta' \partial_{a_1})F} \chi^{(1), \theta'}(z)\nonumber, \\
\end{eqnarray}
with
\begin{eqnarray}
\mathcal{M}_{\theta\theta'} (a_0, a_1; a_2) &=& \frac{\Gamma(-2\theta' a_1)}{\Gamma(\frac{1}{2} + \theta a_0- \theta' a_1 + a_2)} \frac{\Gamma(1+2\theta a_0)}{\Gamma(\frac{1}{2} + \theta a_0- \theta' a_1 - a_2)}.
\end{eqnarray}
Here $\theta=\pm$ are the two fusion channels, $a$ is defined by the Matone relation \cite{Matone:1995rx}
\begin{eqnarray}
\label{eq:matone}
u = -a^2 + a_t^2 -\frac{1}{4} + a_0^2 + t\partial_{t} F,
\end{eqnarray}
and $F(t)$ is the classical conformal block, related to the four-point conformal block as
\begin{eqnarray}
G(\Delta_{0}, \Delta_1, \Delta_{t}, \Delta_{\infty}; \Delta; z_0) = z_0^{\Delta-\Delta_{t}-\Delta_0} e^{b^{-2}(F(t) + \mathcal{O}(b^2))} .
\end{eqnarray}
\subsection{Retarded Green's function for uncharged $AdS_5$ blackhole}
In the previous section, we have discussed how to identify the semiclassical BPZ equation in the two dimensional Liouville theory with the Heun’s equation.  Such a fascinating correlation enables us to write connection formula relating two separate solutions evaluated at two distinct singularities of the Heun's equation \eqref{hypheun} in a seemingly different context.  In particular, we use the connection formula that relates the solution near black hole horizon $(z= t)$ to the solution close to the boundary $(z = 1)$.
\begin{equation} \label{chiexp}
\begin{aligned}
\chi_{\omega \lambda}^{(t), \text { in }}(z)=
 \Big[\Big( & \mathcal{M}_{-+}\left(a_t, a ; a_0\right) \mathcal{M}_{-+}\left(a, a_1 ; a_{\infty}\right) t^{ a} e^{-\frac{1}{2} \partial_a F}   \\
& + \mathcal{M}_{--}\left(a_t, a ; a_0\right) \mathcal{M}_{++}\left(a, a_1 ; a_{\infty}\right) t^{-a} e^{\frac{1}{2} \partial_a F}  \Big) e^{-\frac{1}{2}\partial_{a_1} F}  \chi_{\omega \lambda}^{(1), +}(z) \\
& + \Big( \mathcal{M}_{-+}\left(a_t, a ; a_0\right) \mathcal{M}_{--}\left(a, a_1 ; a_{\infty}\right) t^{ a} e^{-\frac{1}{2} \partial_a F}   \\ & + \mathcal{M}_{--}\left(a_t, a ; a_0\right) \mathcal{M}_{+-}\left(a, a_1 ; a_{\infty}\right) t^{-a} e^{\frac{1}{2} \partial_a F} \Big) e^{\frac{1}{2}\partial_{a_1} F}  \chi_{\omega \lambda}^{(1), -}(z) \Big)\Big] \\ &  t^{\frac{1}{2}-a_0-a_t}(1-t)^{a_t-a_1}  e^{-\frac{1}{2}\partial_{a_t} F},
\end{aligned}
\end{equation}
where $\mathcal{M}$'s are given by \\
\begin{equation}\label{hyppara}
\begin{array}{|c|c|}
\hline \mathcal{M}_{-+}\left(a_t, a ; a_0\right)  &  \frac{\Gamma\left(-2 a \right)}{\Gamma\left(\frac{1}{2}-a_t- a +a_0\right)} \frac{\Gamma\left(1-2  a_t\right)}{\Gamma\left(\frac{1}{2}-a_t-a-a_0\right)}, \\
\hline \mathcal{M}_{-+}\left(a, a_1 ; a_{\infty}\right) & \frac{\Gamma\left(-2 a_1 \right)}{\Gamma\left(\frac{1}{2}-a- a_1 +a_{\infty}\right)} \frac{\Gamma\left(1-2  a\right)}{\Gamma\left(\frac{1}{2}-a-a_1-a_{\infty}\right)},\\
\hline \mathcal{M}_{--}\left(a_t, a ; a_0\right)   & \frac{\Gamma\left(2 a\right)}{\Gamma\left(\frac{1}{2}-a_t+a+a_0\right)} \frac{\Gamma\left(1-2  a_t\right)}{\Gamma\left(\frac{1}{2}-a_t+a-a_0\right)},
\\
\hline \mathcal{M}_{++}\left(a, a_1 ; a_{\infty}\right) & \frac{\Gamma\left(-2  a_1\right)}{\Gamma\left(\frac{1}{2}+ a-a_1+a_{\infty}\right)} \frac{\Gamma\left(1+2 a\right)}{\Gamma\left(\frac{1}{2}+a-a_1-a_{\infty}\right)}, \\
\hline \mathcal{M}_{--}\left(a, a_1 ; a_{\infty}\right) & \frac{\Gamma\left(2 a_1\right)}{\Gamma\left(\frac{1}{2}-a+a_1+a_{\infty}\right)} \frac{\Gamma\left(1-2  a \right)}{\Gamma\left(\frac{1}{2}-a+a_1-a_
{\infty}\right)},\\ 
\hline \mathcal{M}_{+-}\left(a, a_1 ; a_{\infty}\right)  & \frac{\Gamma\left(2 a_1\right)}{\Gamma\left(\frac{1}{2}+ a+a_1+a_{\infty}\right)} \frac{\Gamma\left(1+2 a\right)}{\Gamma\left(\frac{1}{2}+a+a_1-a_{\infty}\right)}.\\ 
\hline
\end{array}
\end{equation}

$G_R(\omega, \lambda)$ is the ratio of the coefficient of the normalizable mode to that of the non-normalizable mode of the asymptotic field expansion near the boundary. We see that right hand side of expression \eqref{chiexp} is the asymptotic field expansion near the boundary having $\chi_{\omega \lambda}^{(1),+}(z)$ and $\chi_{\omega \lambda}^{(1),-}(z)$ as the normalizable and the non-normalizable modes respectively.

Therefore we get, 
\begin{small}
\begin{eqnarray}
&&
G_R(\omega,\lambda)\nonumber \\
&&  = \left(r_{h}^2 -{L^2}\right)^{2 a_1} e^{-\partial_{a_1} F} 
\frac{\sum_{\sigma^{\prime}= \pm} \mathcal{M}_{-\sigma^{\prime}}\left(a_t, a ; a_0\right) \mathcal{M}_{\left(-\sigma^{\prime}\right)+}\left(a, a_1 ; a_{\infty}\right) t^{\sigma^{\prime} a} e^{-\frac{\sigma^{\prime}}{2} \partial_a F}}{\sum_{\sigma= \pm} \mathcal{M}_{-\sigma}\left(a_t, a ; a_0\right) \mathcal{M}_{(-\sigma)-}\left(a, a_1 ; a_{\infty}\right) t^{\sigma a} e^{-\frac{\sigma}{2} \partial_a F}} \nonumber \\ 
\end{eqnarray}
\end{small}
With the knowledge of $F(t)$ and the information provided in \eqref{hyppara} and \eqref{u}  we achieve the exact form of retarded Green's function for $AdS_{4+1}$ hyperbolic blackhole spacetime. 
\subsection{Analytic map between spherically symmetric and hyperbolic $AdS$ black holes} \label{mapping1}

It is interesting to note that a set of complex mappings defined over the coordinates of AdS hyperbolic blackhole reproduces the metric of AdS Schwarzschild blackhole. In particular, if we make use of the following complex mappings, 
\begin{eqnarray} \label{eqn30}
    \chi \rightarrow -i \psi, ~~ r\rightarrow i \hat{r} ~~\text{and}~~ t \rightarrow \hat{t}
\end{eqnarray}
then the metric of AdS hyperbolic blackhole takes the form as, 
\begin{eqnarray}
\label{spherical}
    ds^2=-\left( 1-\frac{M}{\hat{r}^2}+\frac{\hat{r}^2}{L^2}\right)d\hat{t}^2+\frac{d\hat{r}^2}{\left( 1-\frac{M}{\hat{r}^2}+\frac{\hat{r}^2}{L^2}\right)}+\hat{r}^2d\Omega^2_3
\end{eqnarray}
We identify \eqref{spherical} as the metric of an AdS Schwarzschild blackhole with the horizon located at 
\begin{equation}
\hat{r}_h = \sqrt{\frac{\sqrt{L^4 + 4 M L^2} - L^2}{2}}
\label{sphhor}
\end{equation}

As a consequence of this complex mapping we observe, 
\begin{eqnarray}
    T = \frac{1}{\pi}\left(\frac{r_h}{L^2}-\frac{1}{2r_h}\right) \rightarrow ~~ i ~\frac{1}{\pi}\left(\frac{\hat{r}_h}{L^2}+\frac{1}{2\hat{r}_h}\right)  = i \hat{T},
\end{eqnarray}
where $T$ and $\hat{T}$  are the Hawking temperatures of the AdS hyperbolic blackhole and the AdS Schwarzschild blackhole respectively.

It is interesting to notice that in addition to $\chi \rightarrow -i \psi, t\rightarrow i \hat{t}$ and $r\rightarrow i \hat{r}$, if we impose a specific  set of changes of the parameters of AdS hyperbolic spacetime, 
\begin{eqnarray}
\label{eqn301}
    \omega\rightarrow i \hat{\omega},~~ \lambda^2+1\rightarrow-l(l+2),
\end{eqnarray} 
we can reproduce the parameters of Heun's equation for AdS Schwarzschild blackhole \cite{Dodelson:2022yvn}. 
\begin{equation}
\label{trans1}
\begin{split}
t = \frac{r_{h}^2}{2 r_{h}^2-L^2}  & \rightarrow \frac{\hat{r}_{h}^2}{2 \hat{r}_{h}^2+L^2} = \hat{t}, \\
a_0 = 0 & \rightarrow  0 =\hat{a}_0,  \\
a_t = \frac{i \omega}{2} \frac{r_{h}L^2}{2 r_{h}^2-L^2}  & \rightarrow   \frac{i \hat{\omega}}{2} \frac{\hat{r}_{h}L^2}{2 \hat{r}_{h}^2+L^2} =\hat{a}_t, \\
a_1 = \frac{\Delta-2}{2}  & \rightarrow   \frac{\Delta-2}{2}  =\hat{a}_1, \\
a_{\infty} = \frac{\omega L^2}{2} \frac{\sqrt{r_{h}^2-L^2}}{2 r_{h}^2-L^2}   & \rightarrow   \frac{\hat{\omega} L^2}{2} \frac{\sqrt{\hat{r}_{h}^2+L^2}}{2 \hat{r}_{h}^2+L^2}  =\hat{a}_{\infty},
\end{split}
\end{equation}
and
\begin{equation}
\label{transform}
\begin{split}
     & u =  -\frac{(\lambda^{2} +1)L^2+2\left(2 r_{h}^2-L^2\right)+r_{h}^2 \Delta(\Delta-4)}{4\left(r_{h}^2-L^2\right)} 
 +\frac{r_{h}^2 L^4}{(r_{h}^2-L^2)} \frac{\omega^2}{4\left(2 r_{h}^2-L^2\right)} \\
  & \rightarrow  \hat{u}=-\frac{l(l+2)L^2+2(2\hat{r}_h^2+L^2)+\hat{r}_h^2\Delta(\Delta-4)}{4(\hat{r}_h^2+L^2)}+\frac{L^4\hat{r}_h^2}{L^2+\hat{r}_h^2}\frac{\hat{\omega
    }^2}{4(2\hat{r}_h^2+L^2)}. 
\end{split}
\end{equation}
Equations \eqref{trans1} and \eqref{transform} provide a precise recipe for reproducing the retarded Green's function in AdS Schwarzschild blackhole spacetime from AdS hyperbolic blackhole spacetime and vice versa.  

\section{Holographic computation of exact retarded Green's function for thermal CFT with chemical potential in $R^1 \times H^3$}
\label{sec:hypbh1}
In this section, we explicitly compute the exact form of retarded Green's function for a four-dimensional thermal CFT with chemical potential having a well-defined gravity dual known as charged $AdS_5$ hyperbolic black hole. 
The five-dimensional Einstein-Maxwell theory with a negative cosmological constant is described by the following action 
\begin{equation}
    S=\frac{1}{16\pi G_5}\int d^5x \sqrt{-g}\left(R+\frac{12}{L^2}-\frac{1}{g_s^2} F_{\mu \nu}F^{\mu \nu}\right),
\end{equation}
Solutions to the equation of motion deduced from the above action yield the metric for charged hyperbolic $AdS_{4+1}$ black hole,
\begin{equation}
d s_{HAdS_5}^2=-f(r) d t^2+\frac{d r^2}{f(r)}+r^2 d H_3^2; \quad \quad A=\mu \left(1-\frac{r_+^2}{r^2}\right) dt
\end{equation}
Here gauge field $A_{\nu}$ is chosen in such a way that it is zero at the horizon $r_+$. The related blackening factor $f(r)$ and the chemical potential $\mu$ \cite{Svesko:2020dfw} are given as,
\begin{equation}
 f(r)=-1-\frac{M}{r^2}+\frac{r^2}{L^2}+ \frac{Q^2}{r^4}; \quad \text{and} \quad  \mu = \sqrt{\frac{3}{4}} \frac{g_s Q}{r_+^2}
\end{equation}
Here, $r_+$ is the outer horizon and the  black hole parameters $M$ and $Q$ are related to the black hole ADM mass and blackhole charge in the following way \cite{Huang:2022avq},
\begin{eqnarray}
    M_{ADM}= \frac{3 M \Sigma_3}{16 \pi G_5} \quad \text{and} \quad Q_e = \frac{\sqrt{3} \Sigma_3 Q}{4 \pi g_s G_5}
\end{eqnarray}
where $\Sigma_3$ is the volume of the 3-dimensional unit hyperboloid. For convenience, we re-express the blackening factor $f(r)$  as follows
\begin{equation}
f(r)=\frac{(r^2-r_+^2)(r^2-r_-^2)(r^2-r_0^2)}{L^2 r^4}
\end{equation}
where $r_-$ and $r_0$ are given in terms of $r_+$ , $L$ and $Q$ as
\begin{equation}
    r_-^2 = \frac{L^2}{2} \left( 1-\frac{r_+^2}{L^2}+\sqrt{\frac{4Q^2}{r_+^2 L^2}+\left(1-\frac{r_+^2}{L^2}\right)^2}\right),
\end{equation}
\begin{equation}
     r_0^2 = \frac{L^2}{2} \left( 1-\frac{r_+^2}{L^2}-\sqrt{\frac{4Q^2}{r_+^2 L^2}+\left(1-\frac{r_+^2}{L^2}\right)^2}\right).
\end{equation}
and they satisfy the following constraint, 
\begin{equation}
    r_0^2+r_+^2+r_-^2=L^2.
\end{equation}
One can interpret $r_-$ as the inner horizon, however, $r_0$ has no physics interpretation since it is a complex-valued quantity. The Hawking temperature at the outer horizon of the hyperbolic AdS blackhole can be evaluated as \cite{He:2010zb,Huang:2022avq}, 
\begin{eqnarray}
    T^c = \frac{1}{\pi}\Big(\frac{r_+}{L^2}-\frac{Q^2}{2 r_+^5}+\frac{1}{2 r_+} \Big)
\end{eqnarray}
The Klein-Gordon equation, satisfied by a charged scalar field $\phi^c$ of mass $m$ in the charged hyperbolic AdS blackhole is 
\begin{equation}\label{KGeqchads}
\frac{1}{\sqrt{-g}}D_{\mu}\left(\sqrt{-g}g^{\mu\nu}D_{\nu}\right)\phi^c-m^2 \phi^c=0,
\end{equation}
where $D_{\mu}=\nabla _{\mu}-i e A_{\mu}$ is the covariant derivative . \\
Decomposing the scalar field $\phi^c(t,r,\chi,\Omega_2)$ by using the Fourier modes, we find
\begin{equation}\label{chargefourier}
\phi^c(t,r,\chi,\Omega_2)=\int d \omega \int d \lambda \sum_{\Vec\mu} e^{-i \omega t} \Psi^c_{\omega \lambda }(r) Y_{\lambda {\Vec\mu}}\left(\chi, \Omega_2\right)
\end{equation}
where $ Y_{\lambda {\Vec\mu}}\left(\chi, \Omega_2\right)$ is the hyperbolic spherical harmonic previously introduced in \eqref{hyperharmonic}.
Inserting  \eqref{chargefourier} into \eqref{KGeqchads}, the radial equation takes the form as,
\begin{equation}
\left(\frac{1}{r^3} \partial_r\left(r^3 f(r) \partial_r\right)+\frac{1}{f(r)}\left(\omega+\frac{\sqrt{3}eQ}{2r^2}\right)^2-\frac{\left(\lambda^2+1\right)}{r^2}-\frac{\Delta(\Delta-4)}{L^2}\right) \Psi^c_{\omega \lambda}(r)=0
\end{equation}
Imposing the ingoing boundary condition at the horizon, we find
\begin{equation}
{\Psi^c_{\omega \lambda}}^{\text {in }}(r)=\left(r-r_{h}\right)^{-\frac{i (\omega+\omega_c)r_+^3}{2 (r_+^2-r_-^2)(r_+^2-r_0^2)} }+\ldots
\end{equation}
where $\omega_c=\frac{\sqrt{3}rQ}{2r_+^2}$.\\
In order to compute Heun's equation for the charged AdS hyperbolic blackhole, we employ the following transformations,
\begin{equation}
s  =\frac{r^2-r_-^2}{r^2-L^2+r_+^2+r_+^2} 
\end{equation}
\begin{equation}
\Psi^c_{\omega \lambda}(r)  =\left(r^3 f(r) \frac{d s}{d r}\right)^{-1 / 2} X^c_{\omega \lambda}(s)
\end{equation}
In terms of the transformed versions of the radial coordinate and the radial mode, 
we obtain Heun's differential equation in normal form
\begin{align}
\left(\partial_{s}^2+\frac{\frac{1}{4}-{(a^c_1)}^2}{(s-1)^2}-\frac{\frac{1}{2}-{(a^c_0)}^2-{(a^c_1)}^2-{(a^c_t)}^2+{(a^c_\infty)}^2+u^c}{s(s-1)}+\frac{\frac{1}{4}-{(a^c_t)}^2}{(s-t^c)^2}+\frac {u^c}{s(s-t^c)}+\frac{\frac{1}{4}-{(a^c_0)}^2}{{s}^2}\right)\chi^c_{\omega\ell}(s)=0.
\label{hypheunch}
\end{align}   
where the horizon and the boundary are located at $s=t^c$ and at $s=1$ respectively. The parameters of Heun's equation can be written in terms of the black hole parameters as follows

\begin{equation}\label{chargehyppara1}
t^c=\frac{r_+^2-r_-^2}{2r_+^2+r_-^2-L^2},
\end{equation}
\begin{equation}\label{chargehyppara2}
a^c_0=\frac{i L^2 r_- (r_+^2 \omega_c+r_-^2 \omega)}{2(r_+^2-r_-^2)(r_+^2+2r_-^2-L^2)},
\end{equation}
\begin{equation}\label{chargehyppara3}
a^c_t=\frac{i L^2 r_+^3 (\omega+ \omega_c)}{2(r_+^2-r_-^2)(2r_+^2+r_-^2-L^2)},
\end{equation}
\begin{equation}\label{chargehyppara4}
a^c_1=\frac{\Delta-2}{2}
\end{equation}
\begin{equation}\label{chargehyppara5}
a^c_\infty=\frac{L^2 \sqrt{r_+^2+r_-^2-L^2}\left(\omega(r_+^2+r_-^2-L^2)-\omega_c r_+^2\right)}{2(2r_+^2+r_-^2-L^2)(r_+^2+2r_-^2-L^2)},
\end{equation}
and
\begin{equation} \label{uexp}
\begin{aligned}
u^c = & -\frac{(\lambda^{2} +1)L^2+r_+^2 \Delta(\Delta-4)+2(2r_+^2+r_-^2-L^2)}{4(r_+^2+2r_-^2-L^2)}  +\frac{L^4 r_+^2 (\omega+ \omega_c) \left(r_+^2(\omega-\omega_c)-r_-^2(3\omega+\omega_c)\right)}{(r_+^2-r_-^2)^2(2r_+^2+r_-^2-L^2)(r_+^2+2r_-^2-L^2)}.
\end{aligned}
\end{equation}
Finally, the retarded Green's function can be computed as,
\begin{eqnarray}
&&
G^c_R(\omega,\lambda) = \Big(r_{+}^2 - 
 2 r_{-}^2 - {L^2}\Big)^{2 a^c_1} e^{-\partial_{a^c_1} F} \times \nonumber \\
&& \frac{\sum_{\sigma^{\prime}= \pm} \mathcal{M}_{-\sigma^{\prime}}\left(a^c_t, a^c ; a^c_0\right) \mathcal{M}_{\left(-\sigma^{\prime}\right)+}\left(a^c, a^c_1 ; a^c_{\infty}\right) {(t^c)}^{\sigma^{\prime} \cdot a^c} e^{-\frac{\sigma^{\prime}}{2} \partial_{a^c} F}}{\sum_{\sigma= \pm} \mathcal{M}_{-\sigma}\left(a^c_t, a^c ; a^c_0\right) \mathcal{M}_{(-\sigma)-}\left(a^c, a^c_1 ; a^c_{\infty}\right) {(t^c)}^{\sigma \cdot a^c} e^{-\frac{\sigma}{2} \partial_{a^c} F}} \nonumber, \\ 
\end{eqnarray}

\section{Exact form of retarded Green's function for  $AdS_{5}$ black brane}\label{sec:bb}
In this section, we highlight some important aspects of the retarded Green's function for a thermal CFT in $R^1\times R^3$ dual to the $AdS_5$ black brane both from intrinsic and limiting perspectives. 
Here we use the radial coordinate $r$ and also the inverse radial coordinate $z = \frac{L^2}{r}$ and show a curious relation between the parameters of Heun's equation expressed in both coordinate systems.
\subsection{Intrinsic analysis}
$AdS_5$ black brane metric in the Schwarzschild coordinates is given by 
\begin{eqnarray}
\label{flat1}
d s^2=-f(\hat{r}) d t^2+\frac{d \hat{r}^2}{f(\hat{r})}+\frac{\hat{r}^2}{L^2} d {\Vec{x}}_3^2
\end{eqnarray}
where the blackening factor $f(\hat{r})$ and spacial part of boundary metric is 
\begin{equation}
\label{flat2}
 f(\hat{r})= -\frac{M}{\hat{r}^2} + \frac{\hat{r}^2}{L^2} = \frac{\hat{r}^2}{L^2} \left(1-\frac{\hat{r}_b^4}{\hat{r}^4}\right),
\end{equation}
where $r_b = L^{1/2} M^{1/4}$ is the location of the horizon and $T^b = \frac{\hat{r}_b}{\pi L^2}$ is the Hawking temperature.

Like the previous cases, here we consider scalar field $\phi^b$ of mass $m$, satisfying 
the Klein-Gordon equation
\begin{equation}\label{KGflat}
\frac{1}{\sqrt{-g}}\partial_{\mu}\left(\sqrt{-g}g^{\mu\nu}\partial_{\nu}\right)\phi^b-m^2 \phi^b=0,
\end{equation}
where $g_{\mu \nu}$ is given by
\eqref{flat1} and \eqref{flat2}. The Fourier decomposition of the scalar field is
\begin{equation}\label{bbfourier}
\phi^b(t,\hat{r},\Vec{x})=\int d \hat{\omega} \int d^3 k  \ e^{-i \hat{\omega} t} e^{-i \Vec{k} . \Vec{x}} {\Psi^b}_{\hat{\omega} \Vec{k}}(\hat{r})
\end{equation}
Inserting  \eqref{bbfourier} in to \eqref{KGflat}, the radial equation reads
\begin{equation} \label{bbeqn}
\left(\frac{1}{\hat{r}^3} \partial_{\hat{r}}\left(\hat{r}^3 f(\hat{r}) \partial_{\hat{r}}\right)+\frac{{\hat{\omega}}^2}{f(\hat{r})}-|\Vec{k}|^2 \frac{L^2}{\hat{r}^2}-\frac{\Delta(\Delta-4)}{L^2}\right) {\Psi^b}_{\hat{\omega} \Vec{k}}(\hat{r})=0
\end{equation}
Under the following coordinate transformation 
\begin{equation}
\label{bby}
    y=\frac{\hat{r}^2}{\hat{r}^2+ \hat{r}_b^2}
\end{equation}
and with the redefined field transformation,
\begin{equation}
\label{redefbb}
    \Psi^{b}_{\hat{\omega} \Vec{k}}(\hat{r})  =\left(\hat{r}^3 f(\hat{r}) \frac{d y}{d \hat{r}}\right)^{-1 / 2} {X^b}_{\hat{\omega} \Vec{k}}(y)
\end{equation}
 we find the radial equation \eqref{bbeqn} reduces into the normal form of Heun's differential equation given by
 \begin{equation}
\left(\partial_y^2+\frac{\frac{1}{4}-{(a^b_1)}^2}{(y-1)^2}-\frac{\frac{1}{2}-{(a^b_0)}^2-{(a^b_1)}^2-{(a^b_t)}^2+{(a^b_\infty)}^2+u^b}{y(y-1)}+\frac{\frac{1}{4}-{(a^b_t)}^2}{(y-t^b)^2}+\frac{u^b}{y(y-t^b)}+\frac{\frac{1}{4}-{(a^b_0)}^2}{y^2}\right){X^b}_{\hat{\omega}\Vec{k}}(y)=0.
\label{flatheun}
\end{equation}
 
Moreover, we can express the parameters of Heun's equation in terms of the parameters of the $AdS_5$ blackbrane, 
\begin{equation} \label{rpara}
\begin{array}{|c|c|c|c|c|c|}
\hline \text { Heun's parameter } & t^b & a^b_0 & a^b_t & a^b_1 & a^b_{\infty} \\
\hline \text { AdS black brane parameter} & \frac{1}{2} & 0 & \frac{i \check{\omega}}{4\pi} & \frac{\Delta-2}{2} & \frac{\check{\omega}}{4 \pi}  \\
\hline
\end{array}
\end{equation}
and $u^b$ is given by 
\begin{equation} \label{urexp}
    u^b = \frac{\check{\omega}^2 -2 |\Vec{\tilde{k}}|^2} {8 \pi^2} - \frac{1}{4} (\Delta -2)^2
\end{equation}
Here we have introduced  a set of dimensionless parameters $|\Vec{\tilde{k}}|= \frac{|\Vec{k}|}{ T^b}$ and $\check{\omega} = \frac{\hat{\omega}}{T^b}$. 

$AdS_5$ black brane metric in terms of the inverse radial coordinate $\hat{z} = \frac{L^2}{\hat{r}}$  is given as 
\begin{eqnarray}
d s^2=\frac{L^2}{\hat{z}^2}\left[-\left(1-\frac{\hat{z}_b^4}{\hat{z}^4}\right) d t^2+\frac{d \hat{z}^2}{\left(1-\frac{\hat{z}_b^4}{\hat{z}^4}\right)}+d x_1^2+d x_2^2+d x_3^2 \right]
\end{eqnarray}
Similar to \eqref{bbeqn}, the radial mode satisfies, 
\begin{equation} \label{bbzeqn}
\left(\hat{z}^3 \partial_{\hat{z}}\left(\frac{1}{\hat{z}^3} f(\hat{z}) \partial_{\hat{z}}\right)+\frac{\hat{\omega}^2}{f(\hat{z})}-|\Vec{k}|^2 -\frac{\Delta(\Delta-4)}{\hat{z}^2}\right) {\Psi^b}_{\hat{\omega} \Vec{k}}(\hat{z})=0
\end{equation}
By the virtue of following transformations
\begin{eqnarray}
\tilde{y}=\frac{\hat{z}^2}{\hat{z}^2+ \hat{z}_b^2}, ~~ {\Psi^b}_{\hat{\omega} \Vec{k}}(\hat{z})  =\left(\frac{1}{\hat{z}^3} f(\hat{z}) \frac{d \tilde{y}}{d \hat{z}}\right)^{-1 / 2} {X^b}_{\hat{\omega} \Vec{k}}(\tilde{y}),
\end{eqnarray}
the radial equation \eqref{bbzeqn} reduces into the normal form of Heun's differential equation that takes  exactly the same form as \eqref{flatheun} except $y$ is now replaced by $\tilde{y}$. Note that in terms of $y$ coordinate, the horizon and boundary of the AdS black brane is mapped to $y=\frac{1}{2}$ and $y=1$ respectively while in $\tilde{y}$ coordinate, those are located as  $\tilde{y}=\frac{1}{2}$ and $\tilde{y}=0$ respectively. 
Most importantly, the Heun's parameters are related to the  AdS black brane parameters as,
\begin{eqnarray} \label{zpara}
\begin{array}{|c|c|c|c|c|c|}
\hline  &  &  &  &  &   \\
 \text { Heun's parameter } & \tilde{t^b} & \tilde{a^b_0} & \tilde{a^b_t} & \tilde{a^b_1} & \tilde{a^b_\infty} \\
\hline  &  &  &  &   &  \\  
\text { AdS black brane parameter } & \frac{1}{2} & \frac{\Delta-2}{2} & \frac{i \check{\omega}} {4 \pi} & 0 & \frac{\check{\omega}}{4 \pi}  \\
\hline
\end{array}
\end{eqnarray}
and $\tilde{u^b}$ is given by 
\begin{equation} \label{uzexp}
    \tilde{u^b} =- \frac{\check{\omega}^2 -2 |\Vec{\tilde{k}}|^2} {8 \pi^2} + \frac{1}{4} (\Delta -2)^2
\end{equation} 
  On comparing Heun's parameters in \eqref{rpara} and \eqref{zpara}, we notice that the parameters $a_0$ and $a_1$ are interchanged.   Besides this, the parameter $u$ in \eqref{urexp} and \eqref{uzexp} admits a relative minus sign.
  Such changes can easily be regarded as necessary by performing the appropriate coordinate transformation $y = 1-\tilde{y}$ in \eqref{flatheun} to re-express it in terms of $\tilde{y}$ coordinate. 
 
\subsection{Limiting perspective}

In \cite{Dodelson:2022yvn}, the authors have demonstrated how to derive the Heun's parameters of an $AdS_5$ blackbrane by taking the high-temperature limit of those of the $AdS_5$ Schwarzschild blackhole. Let's recall Heun's parameters of the $AdS_5$ Schwarzschild blackhole as listed below,
\begin{equation}
\label{planner}
\begin{array}{|c|c|c|c|c|c|}
\hline \text { Heun's parameter } & \hat{t} & \hat{a_0} & \hat{a_t} & \hat{a_1} & \hat{a}_{\infty} \\
\hline \text {$AdS_5$ Schwarzschild black hole } & \frac{\hat{r}_{h}^2}{2 \hat{r}_{h}^2+L^2} & 0 & \frac{i \hat{\omega}}{2} \frac{\hat{r}_{h}L^2}{2 \hat{r}_{h}^2+L^2} & \frac{\Delta-2}{2} & \frac{\hat{\omega} L^2}{2} \frac{\sqrt{\hat{r}_{h}^2+L^2}}{2 \hat{r}_{h}^2+L^2}  \\
\hline
\end{array}
\end{equation}
Note that the high-temperature limit of the $AdS_5$ Schwarzschild blackhole can be realized by a large $M$ limit. Consequently, once we implement $M \to \infty$ in equation \eqref{sphhor} we obtain, 
\begin{equation}
\hat{r}_h = \hat{r}_b = L^{1/2} M^{1/4} 
\end{equation}
Moreover, the same limit leads us to  recover the temperature of the $AdS_5$ blackbrane,
\begin{eqnarray}
\lim _{M \to \infty} \hat{T} = \hat{T}_{\infty}= {T}^{b}= \frac{\hat{r}_b}{L^2\pi}
\label{sphbbT}
\end{eqnarray}
Further, with $\hat{\omega}\rightarrow\infty $ and $l\rightarrow\infty$ and by keeping the ratios $\frac{\hat{\omega}}{\hat{T}_{\infty}} = \check{\omega}$ and $\frac{l}{L \hat{T}_{\infty}} = |\Vec{\tilde{k}}|$ finite, one can recover the Heun's parameters for the $AdS_5$ blackbrane \cite{Dodelson:2022yvn}, 
\begin{equation}\label{plannersch}
\begin{array}{|c|c|c|c|c|c|}
\hline \text { Heun's parameter } & {t}^{b} & {a}^{b}_0 & {a}^{b}_t & {a}^{b}_1 & {a}^{b}_{\infty} \\
\hline \text {$AdS_5$ black brane} & \frac{1}{2} & 0 & \frac{i \check{\omega}}{4\pi} & \frac{\Delta-2}{2} & \frac{\check{\omega} }{4 \pi} \\
\hline
\end{array}
\end{equation}
along with
\begin{eqnarray}
    u^{b}=\frac{\check{\omega}^2 -2{|\Vec{\tilde{k}}|}^2}{8\pi^2 }-\frac{1}{4}(\Delta-2)^2.
\end{eqnarray}

Similarly, here we show that the same result that we have obtained for $AdS_5$ blackbrane can be recovered by taking a high-temperature limit on the $AdS_5$ hyperbolic blackhole followed by a set of complex mappings. 
Like the $AdS_5$ Schwarzschild case, here the high-temperature limit is implemented also as the large $M$ limit. By comparing \eqref{hyphor}  with  $\hat{r}_b$ for large values of $M$, we conclude ${r}_h = \hat{r}_b = L^{1/2} M^{1/4} $. Moreover, we recover the temperature of $AdS_5$ blackbrane as, 
\begin{eqnarray}
\lim _{M \to \infty} T = T_{\infty}= {T}^{b}= \frac{\hat{r}_b}{L^2\pi}
\label{hypbbT}
\end{eqnarray}
If we evoke ${\omega}\rightarrow\infty $ and $\lambda \rightarrow\infty$ by keeping the ratios $\frac{{\omega}}{{T}_{\infty}}$ and $\frac{\lambda}{{T}_{\infty}}$ finite, the Heun's parameter of the $AdS_5$ hyperbolic blackhole take the following form,
\begin{eqnarray}
t= \frac{1}{2},  ~~a_0 = 0, ~~ a_t =  \frac{i \omega}{4 \pi T_{\infty}}, ~~a_1 = \frac{\Delta-2}{2}, ~~ a_{\infty} = \frac{\omega L^2}{4 \pi T_{\infty}},~~ u =  \frac{{\omega}^2 L^2-2\lambda^2} {8\pi^2 T_{\infty}^2 L^2}-\frac{1}{4}(\Delta-2)^2
\label{hlimit}
\end{eqnarray}
Now, once we impose a set of complex mappings on the limiting value of the Heun's parameters listed in \eqref{hlimit},
\begin{eqnarray}
 ~~\omega \to i \hat{\omega}, ~~ T_{\infty} \to i T_{\infty}, ~~ \lambda \to i l 
\end{eqnarray}
we recover the Heun's parameters for $AdS_5$ blackbrane. Note that $\lambda \to i l$ is a remnant of the transformation $\lambda^2+1\rightarrow-l(l+2)$ when $\lambda$ and $l$ are very large. 
\section{Exact thermal correlators for Rindler $ AdS_5 $}\label{sec:rindler}
In the realm of $AdS$ space, an observer moving with an acceleration beyond a critical value 
experiences a Rindler horizon. In particular, a finite sub-region in the global $AdS_5$ spacetime bounded with a finite interval of global $AdS_5$ time can be foliated in terms of Rindler-$AdS_5$ coordinates of two casually disconnected wedges separated by the Rindler horizon. The boundary of the individual wedge is described as $R\times H^3$ spacetime. The dual $CFT_4$ living in the boundary of one of the Rindler $AdS_5$ wedges is thermal in nature where the corresponding temperature is holographically dual to the Hawking temperature at Rindler $AdS_5$ horizon. Rindler $AdS$ metric in 5-dimension is given by \cite{Parikh:2012kg, Ahn:2020csv}
\begin{equation}
d s_{RAdS_5}^2=-f(r) d t^2+\frac{d r^2}{f(r)}+r^2 d H_3^2
\end{equation}
where the blackening factor $f(r)$ is
\begin{equation}
 f(r)=\frac{r^2}{L^2}-1
\end{equation}
and \begin{equation}
 d H_3^2=d \chi^2+\sinh ^2 \chi \ d  \Omega_2^2 
\end{equation}
Note that $-\infty<t<\infty$, $\chi \ge 0$
and the horizon is located at $r=L$.

A massive scalar field satisfying the Klein-Gordon equation in Rindler $AdS_5$ background allows the following mode expansion, 
\begin{equation}\label{rindfourier}
\phi(t,r,\chi,\Omega_2)=\int d \omega \int d \lambda \sum_{\Vec\mu} e^{-i \omega t} \psi_{\omega \lambda }(r) Y_{\lambda {\Vec\mu}}\left(\chi, \Omega_2\right)
\end{equation}
where $ Y_{\lambda {\Vec\mu}}\left(\chi, \Omega_2\right)$ is the hyperbolic spherical harmonic satisfying 
\begin{equation}\label{rindhyperharmonic}
\square_{H_{3}} Y_{\lambda_ {\Vec\mu}}\left(\chi, \Omega_{2}\right)=-\left(\lambda^2 +1\right) Y_{\lambda  {\Vec\mu}}\left(\chi, \Omega_{2}\right)
\end{equation}
Inserting  \eqref{fourier} in to \eqref{KGeq}, with \eqref{hyperharmonic}, the radial equation reads
\begin{equation} \label{reqnforrads}
\left(\frac{1}{r^3} \partial_r\left(r^3 f(r) \partial_r\right)+\frac{\omega^2}{f(r)}-\frac{\left(\lambda^2+1\right)}{r^2}-\frac{\Delta(\Delta-4)}{L^2}\right) \psi_{\omega \lambda}(r)=0
\end{equation}
Redefining the radial coordinate and the radial mode of the field expansion we get,   
\begin{eqnarray}
      z=\frac{r^2}{r^2+n L^2}, ~~~~~\psi_{\omega \lambda}(r)  =\left(r^3 f(r) \frac{d z}{d r}\right)^{-1 / 2} \chi_{\omega \lambda}(z)
\end{eqnarray}
where n is some positive integer.

The radial equation \eqref{reqnforrads} reduces into the normal form of Heun's differential equation given by
\begin{align}
\left(\partial_z^2+\frac{\frac{1}{4}-a_1^2}{(z-1)^2}-\frac{\frac{1}{2}-a_0^2-a_1^2-a_t^2+a_\infty^2+u}{z(z-1)}+\frac{\frac{1}{4}-a_t^2}{(z-t)^2}+\frac{u}{z(z-t)}+\frac{\frac{1}{4}-a_0^2}{z^2}\right)\chi_{\omega\ell}(z)=0.
\label{rindlerheun}
\end{align}

Now the locations of the horizon and the boundary of $RAdS_5$ are mapped to $z=t=\frac{1}{n+1}$ and $z=1$ respectively. Moreover, Heun's parameters are related to the Rinlder $AdS_5$ parameters in the following way,
\begin{equation}
\begin{array}{|c|c|c|c|c|c|}
\hline \text { Heun's parameter} & t & a_0 & a_t & a_1 & a_{\infty} \\
\hline \text { Rindler $AdS_5$ } & \frac{1}{n+1} & \frac{i \lambda}{2} & \frac{i \omega L}{2} & \frac{\Delta-2}{2} & \frac{1}{2}  \\
\hline
\end{array}
\label{heunrind}
\end{equation}
and u is given by 
\begin{equation}
    u = -\frac{n \left((\Delta -4) \Delta +\lambda ^2+L^2 \omega ^2+5\right)+(\Delta -4) \Delta +\lambda ^2-L^2 \omega ^2+3}{4 n}.
\end{equation}

The exact formula for the retarded two-point function can be obtained by using relation \eqref{rtgrnfn} and the result is 
\begin{equation}
\begin{aligned}
& G_R(\omega,\lambda)=\left(n L^2\right)^{2 a_1} e^{-\partial_{a_1} F} 
\frac{\sum_{\sigma^{\prime}= \pm} \mathcal{M}_{-\sigma^{\prime}}\left(a_t, a ; a_0\right) \mathcal{M}_{\left(-\sigma^{\prime}\right)+}\left(a, a_1 ; a_{\infty}\right) t^{\sigma^{\prime} a} e^{-\frac{\sigma^{\prime}}{2} \partial_a F}}{\sum_{\sigma= \pm} \mathcal{M}_{-\sigma}\left(a_t, a ; a_0\right) \mathcal{M}_{(-\sigma)-}\left(a, a_1 ; a_{\infty}\right) t^{\sigma a} e^{-\frac{\sigma}{2} \partial_a F}}
\end{aligned}
\end{equation}

\section{Conclusion}\label{sec:conclusion}

In this work, we have  holographically computed the exact form of retarded Green's function for thermal CFT in $R^1 \times H^3$. We generalized our analysis for a thermal CFT in the presence of chemical potential in $R^1 \times H^3$. For both cases, our computation is done holographically in the respective dual gravity backgrounds described by $AdS_5$ hyperbolic blackhole and its charged cousin. It is important to mention that 
our results are well-defined within the large blackhole approximation. Further, we have computed 
the retarded Green's function for $AdS_5$ blackbrane both in an intrinsic way and from a high-temperature limit on the $AdS_5$ hyperbolic blackhole. In the limiting analysis, the high-temperature limit is achieved from $M\to \infty$ followed by a set of complex transformations. Finally, we have also presented an exact form of retarded Green's function for thermal $CFT_4$ dual to the Rindler $AdS_5$ background. Here we find that the final form of Green's function depends on the free integer parameter. We have checked that no such free parameter appears in the results  obtained in the $AdS_5$ hyperbolic blackhole and in its charged version. The holographic interpretation of the free parameter present in the result for Rindler $AdS_5$ requires further investigation. 

It is important to note that in our analysis Zamolodchikov's classical block in the Liouville theory is dubbed as the $F$ function that appears in the exact form of retarded Green's function. However, by virtue of AGT correspondence one can relate this classical block to the conventional $F$ function in $SU(2)$ gauge theory. It is very intriguing to notice that the classical block $F$ is completely fixed by the global part, i.e. $SL(2, \mathbb{R})$ subalgebra of the Virasoro algebra and hence the form of $F$ function is kinematic in nature and does not depend on the details of the theory. The appearance of 2d Louville CFT in the study of higher dimensional thermal CFTs is surprising and needs a deeper analysis to understand the underlying relations among them.

One obvious extension would be to compute retarded Green's function for boosted black brane. Boost in the AdS black brane does not change the singular behaviour of the blackening factor of the solution and hence that of the scalar field equation. We expect that the scalar field equation on the boosted black brane background can be mapped to the Heun equation and it is interesting to see how the parameters of the resulting Heun equation depend upon the boost parameter. 

In the literature, there are closed-form expressions for thermal correlators in the 4-dimensional thermal CFTs using various approximations, such as the geodesic approximation. It is worth checking if we can reproduce those known results by taking the appropriate limit of the exact form of the retarded Green's function presented here. 

One can find the retarded Green's function for currents $\langle J_\mu J_\nu \rangle$ and $ \langle T_{\mu \nu} T_{\rho\sigma}\rangle $ of thermal CFTs in $R^1 \times H^3$ by considering the vector and tensor field equations on the dual black hole background respectively. From $ \langle T_{xx} T_{xx}\rangle $ one can study the pole skipping, hence chaos in such theories.

It will be exciting if these techniques can be generalized for higher dimensional thermal CFTs of both the types $R^1 \times H^{d-1}$ and $R^1 \times S^{d-1}$.

\acknowledgments
The work of AB is supported by the South African Research Chairs Initiative of the Department of Science and Innovation and the National Research Foundation, grant number 78554. The work of TM is supported by the grant SB/SJF/2019-20/08. Arpit Maurya would like to thank the Council of Scientific and Industrial Research (CSIR), Government of India, for the financial support through a research fellowship (File No.: 09/1005(0034)/2020-EMR-I).  

\bibliographystyle{JHEP}
\bibliography{thermal}

\end{document}